\documentclass[11pt, leqno]{article}

\usepackage{amsmath}
\usepackage{amsthm}
\usepackage{amssymb}
\usepackage[dvips]{graphicx,color}

\theoremstyle{plain}
\newtheorem{theorem}{Theorem}[section]
\newtheorem{lemma}[theorem]{Lemma}

\theoremstyle{definition}

\theoremstyle{remark}
\newtheorem{remark}[theorem]{Remark}

\numberwithin{equation}{section}

\begin{document}

\title{\textbf{Another operator-theoretical proof for \\the second-order phase transition in \\the BCS-Bogoliubov model of superconductivity}}
\author{Shuji Watanabe\\
Division of Mathematical Sciences\\
Graduate School of Engineering, Gunma University\\
4-2 Aramaki-machi, Maebashi 371-8510, Japan\\
Email: shuwatanabe@gunma-u.ac.jp}

\date{}

\maketitle

\begin{abstract}
%%%%%%%%%%%%%%%%%%%%%%%%%%%%%%%%%%%
In the preceding papers, imposing certain complicated and strong conditions, the present author showed that the solution to the BCS-Bogoliubov gap equation in superconductivity is twice differentiable only on the neighborhoods of absolute zero temperature and the transition temperature so as to show that the phase transition is of the second order from the viewpoint of operator theory. Instead, we impose a certain simple and weak condition in this paper, and show that there is a unique nonnegative solution and that the solution is indeed twice differentiable on a closed interval from a certain positive temperature to the transition temperature as well as pointing out several properties of the solution. We then give another operator-theoretical proof for the second-order phase transition in the BCS-Bogoliubov model. Since the thermodynamic potential has the squared solution in its form, we deal with the squared BCS-Bogoliubov gap equation. Here, the potential in the BCS-Bogoliubov gap equation is a function and need not be a constant.
%%%%%%%%%%%%%%%%%%%%%%%%%%%%%%%%%%%

\medskip


\medskip

\end{abstract}

%%%%%%%%%%%%%%%%%%%%%%%%%%%%%%%%%%%%%%%%%%%%%%%%%%%%%%%%%%%%%%%%% 1
\section{Introduction}

In the BCS-Bogoliubov model of superconductivity, one does not show that the solution to the BCS-Bogoliubov gap equation is partially differentiable with respect to the absolute temperature $T$. Nevertheless, without such a proof, one partially differentiates the solution and the thermodynamic potential with respect to the temperature twice so as to obtain the entropy and the specific heat at constant volume. One then shows that the phase transition from a normal conducting state to a superconducting state is of the second order. Therefore, if the solution were not partially differentiable with respect to the temperature, then one could not partially differentiate the solution and the thermodynamic potential with respect to the temperature, and hence one could not obtain the entropy and the specific heat at constant volume. As a result, one could not show that the phase transition is of the second order. For this reason, it is highly desirable to show that there is a unique solution to the BCS-Bogoliubov gap equation and that the solution is partially differentiable with respect to the temperature twice.

In the preceding papers (see \cite[Theorems 2.2 and 2.10]{watanabe-seven} and \cite[Theorems 2.3 and 2.4]{watanabe-five}), the present author gave a proof of the existence and uniqueness of the solution and showed that the solution is indeed partially differentiable with respect to the temperature twice on the basis of fixed-point theorems. In this way, the present author gave an operator-theoretical proof of the statement that the phase transition is of the second order, and thus solved the long-standing problem of the second-order phase transition from the viewpoint of operator theory. Here, the potential in the BCS-Bogoliubov gap equation is a function and need not be a constant.

But the present author imposed certain complicated and strong conditions in the preceding papers \cite{watanabe-seven, watanabe-five, watanabe-eight}. As a result, the present author showed that the solution to the BCS-Bogoliubov gap equation is partially differentiable with respect to the temperature only on the neighborhoods of absolute zero temperature $T=0$ and the transition temperature $T=T_c$. Instead, we impose a certain simple and weak condition in this paper. Thanks to this simple and weak condition, we show that there is a unique nonnegative solution and that the solution is indeed partially twice differentiable with respect to the temperature on the interval $[T_0, \, T_c]$ as well as pointing out several properties of the solution. Here, the temperature $T_0$ is defined in \eqref{eqn:tzero} below, and the temperature interval $[T_0,\, T_c]$ can be nearly equal to the whole temperature interval $[0,\, T_c]$ (see Remark \ref{rmk:tzerotc} below).

Differentiating the thermodynamic potential with respect to the temperature, we thus give another operator-theoretical proof for the second-order phase transition. As is well known, the thermodynamic potential has the squared solution in its form, not the solution itself. Therefore, we deal with the squared BCS-Bogoliubov gap equation, not the BCS-Bogoliubov gap equation. From the viewpoint of operator theory, the present author thinks that dealing with the squared BCS-Bogoliubov gap equation provides a straightforward way to show the second-order phase transition.

The BCS-Bogoliubov gap equation \cite{bcs, bogoliubov} is a nonlinear integral equation given by
\begin{equation}\label{eqn:bcseq}
u_0(T,\,x)=\int_I \frac{U(x,\,\xi)\, u_0(T,\, \xi)}{\,\sqrt{\,\xi^2+u_0(T,\, \xi)^2\,}\,}\,
\tanh \frac{\,\sqrt{\,\xi^2+u_0(T,\, \xi)^2\,}\,}{2T}\, d\xi, \   T \geq 0, \   (x, \,\xi) \in I^2.  
\end{equation}
Here, the solution $u_0$ is a function of the absolute temperature $T$ (of a superconductor) and the energy $x$ (of an electron). The closed interval $I$ is given by $I=[\varepsilon,\, \hslash\omega_D]$, where the Debye angular frequency $\omega_D$ is a positive constant and depends on a superconductor, and $\varepsilon>0$ is a cutoff (see the following remark). The potential $U(\cdot,\,\cdot)$ satisfies $U(x,\,\xi)>0$ at all $(x,\,\xi) \in I^2$. Throughout this paper we use the unit where the Boltzmann constant $k_B$ is equal to 1.

\begin{remark}
For simplicity, we introduce the cutoff $\varepsilon>0$ in \eqref{eqn:bcseq}. Here, the cutoff $\varepsilon>0$ is small enough. We see that the cutoff is unphysical, but we introduce it for simplicity.
\end{remark} 

In \cite{odeh, billardfano, vansevenant, bls, chen, deugeihailoss, fhns, fhss, freijihaizlseiringer, hhss, hainzlloss, haizlseiringer, haizlseiringer2, haizlseiringer3, watanabe-one, watanabe-two, watanabe-four, watanabe-seven, watanabe-five, watanabe-eight}, the existence, the uniqueness and several properties of the solution to the BCS-Bogoliubov gap equation were established and studied. See also Kuzemsky \cite[Chapters 26 and 29]{kuzemsky3} and \cite {kuzemsky, kuzemsky2}. Anghel and Nemnes \cite{angnem} and Anghel \cite{ang-one, ang-two} showed that if the physical quantity $\mu$ in the BCS-Bogoliubov model is not equal to the chemical potential, then the phase transition from a normal conducting state to a superconducting state is of the first order under a certain condition without any external magnetic field. Introducing imaginary magnetic field, Kashima \cite{kashima-one, kashima-two, kashima-three, kashima-four} pointed out that the phase transition is of the second-order if and only if a certain value is greater than $\sqrt{17-12\sqrt{2}}$ and that the phase transition is of order $4n+2$ if and only if the value above is less than or equal to $\sqrt{17-12\sqrt{2}}$. Here. $n$ is an arbitrary positive integer.

In this connection, the BCS-Bogoliubov gap equation in superconductivity plays a role similar to that of the Maskawa--Nakajima equation \cite{maskawa-nakajima-one, maskawa-nakajima-two} in elementary particle physics. In Professor Maskawa's Nobel lecture, he stated the reason why he considered the Maskawa-Nakajima equation. See the present author's paper \cite{watanabe-three} for an operator-theoretical treatment of the Maskawa--Nakajima equation.

Squaring both sides of the BCS-Bogoliubov gap equation and putting $f_0(T,\,x)=u_0(T,\,x)^2$ give the squared BCS-Bogoliubov gap equation:
\[
f_0(T,\,x)=\left( \int_I U(x,\,\xi) \sqrt{ \, \frac{f_0(T,\, \xi)}{\, \xi^2+f_0(T,\, \xi)\,} \,} \, \tanh \frac{\,\sqrt{\,\xi^2+f_0(T,\, \xi)\,}\,}{2T}\, d\xi \right)^2.
\]
Let $T_c$ be the transition temperature (see \cite[Definition 2.5]{watanabe-one} for our operator-theoretical definition of $T_c$) and let $D=[T_0,\, T_c] \times I \in \mathbb{R}^2$. Here, $I=[\varepsilon,\, \hslash\omega_D]$. Define our operator $A$ by
\begin{equation}\label{eqn:ourop}
Af(T,\,x)=\left( \int_I U(x,\,\xi) \sqrt{ \, \frac{f(T,\, \xi)}{\, \xi^2+f(T,\, \xi)\,} \,} \, \tanh \frac{\,\sqrt{\,\xi^2+f(T,\, \xi)\,}\,}{2T}\, d\xi \right)^2, \quad (T,\,x) \in D,
\end{equation}
where $f \in W$ (see \eqref{eqn:w} below for the subset $W$). We define our operator $A$ on the subset $W$ and look for a fixed point of our operator $A$. Note that a fixed point of $A$ becomes a solution to the squared BCS-Bogoliubov gap equation, and that its square root becomes a solution to the BCS-Bogoliubov gap equation \eqref{eqn:bcseq}.

Let $U_1$ and $U_2$ be positive constants, where $(0<) \, U_1 \leq U_2$. If the potential $U(\cdot,\,\cdot)$ is a positive constant and $U(x,\,\xi)=U_1$ at all $(x,\,\xi) \in I^2$, then the solution to the BCS-Bogoliubov gap equation \eqref{eqn:bcseq} becomes a function of the temperature $T$ only. Denoting the solution by $T \mapsto \Delta_1(T)$, we have (see \cite{bcs})
\begin{equation}\label{eqn:delta1}
1=U_1 \int_I \frac{1}{\,\sqrt{\,\xi^2+\Delta_1(T)^2\,}\,} \,
 \tanh \frac{\, \sqrt{\,\xi^2+\Delta_1(T)^2\,}\,}{2T}\,d\xi, \quad 0 \leq T \leq\tau_1.
\end{equation}
Here, the temperature $\tau_1>0$ is defined by (see \cite{bcs})
\[
1=U_1 \int_I \frac{1}{\,\xi\,} \, \tanh \frac{\xi}{\,2\tau_1\,}\,d\xi.
\]
The solution $T \mapsto \Delta_1(T)$ is continuous and strictly decreasing with respect to $T$, and moreover, the solution is of class $C^2$ with respect to $T$. For more details, see \cite[Proposition 1.2]{watanabe-one}.

We set $\Delta_1(T)=0$ at all $T \geq \tau_1$. Then \eqref{eqn:delta1} becomes
\[
1>U_1 \int_I \frac{1}{\,\xi\,} \, \tanh \frac{\xi}{\,2T \,}\,d\xi, \quad T>\tau_1.
\]
We choose an arbitrary temperature $T_0 \; (>\tau_1)$. Then, for $T \in [T_0,\, T_c]$,
\begin{equation}\label{eqn:tzero}
U_1 \int_I \frac{1}{\,\xi\,} \, \tanh \frac{\xi}{\,2T \,}\,d\xi <1.
\end{equation}

On the other hand, If $U(x,\,\xi)=U_2$ at all $(x,\,\xi) \in I^2$, then we have the solution $T \mapsto \Delta_2(T)$ to
\begin{equation}\label{eqn:delta2}
1=U_2 \int_I \frac{1}{\,\sqrt{\,\xi^2+\Delta_2(T)^2\,}\,} \,
 \tanh \frac{\, \sqrt{\,\xi^2+\Delta_2(T)^2\,}\,}{2T}\,d\xi, \quad 0 \leq T \leq\tau_2.
\end{equation}
Here, the temperature $\tau_2>0$ is defined by
\[
1=U_2 \int_I \frac{1}{\,\xi\,} \, \tanh \frac{\xi}{\,2\tau_2\,}\,d\xi.
\]
The solution $T \mapsto \Delta_2(T)$ has properties similar to those of the solution $T \mapsto \Delta_1(T)$. We again set $\Delta_2(T)=0$ at all $T \geq \tau_2$.

The inequality $U_1 \leq U_2$ implies
\[
\Delta_1(T) \leq \Delta_2(T) \quad (0 \leq T \leq \tau_2).
\]
For the graphs of $\Delta_1(\cdot)$ and $\Delta_2(\cdot)$, see \cite[Figure 1]{watanabe-five}. 

%%%%%%%%%%%%%%%%%%%%%%%%%%%%%%%%%%%%%%%%%%%%%%%%%%%%%%%%%%%%%%%%%2
\section{Main results}

Suppose that the potential $U(\cdot,\,\cdot)$ in the BCS-Bogoliubov gap equation \eqref{eqn:bcseq} satisfies the following conditions:
\begin{equation}\label{eqn:potential}
U(\cdot,\,\cdot) \in C^2(I^2), \quad (0<) \, U_1 \leq U(x,\,\xi) \leq U_2 \quad \hbox{at all}\quad (x,\,\xi) \in I^2,
\end{equation}
and \eqref{eqn:condition-two} below.

The inequalities $U_1 \leq U(x,\,\xi) \leq U_2$ at all $(x,\,\xi) \in I^2$ imply $\tau_1 \leq T_c \leq \tau_2$ \   (see \cite[Remark 2.6]{watanabe-one}). Set
\[
a=\left\{ \frac{ \displaystyle{ \max_{(x,\,\xi) \in I^2} U(x,\,\xi) }}{\, \displaystyle{ \min_{(x,\,\xi) \in I^2} U(x,\,\xi) } \,} \right\}^2 \   (\geq 1).
\]

\begin{remark}\label{rmk:tzerotc}
The temperatures $\tau_1$, $T_0$, $T_c$, $\tau_2$ satisfy $(0<)\, \tau_1<T_0<T_c \leq \tau_2$. If $U_1$ is small enough, then so is $\tau_1$. Therefore, $T_0$ can be small enough, and hence $T_0$ does not need to be close to the transition temperature $T_c$. As a result, the temperature interval $[T_0,\, T_c]$ can be nearly equal to the whole temperature interval $[0,\, T_c]$. For temperatures at or near zero temperature, the smoothness of the solution to the BCS-Bogoliubov gap equation with respect to such temperatures has been shown in \cite[Theorem 2.2]{watanabe-seven}.  In this paper we thus deal with the temperature interval $[T_0,\, T_c]$.
\end{remark}

Let $W$ be a subset of the Banach space $C(D)$ satisfying
\begin{eqnarray}\label{eqn:w}
W &=& \big\{ f \in C^2(D) (\subset C(D)) :  (0 \leq) \   \Delta_1(T)^2 \leq f(T,\,x) \leq \Delta_2(T)^2, \quad f(T_c,\,x) = 0, \\ \nonumber
& & \frac{f(T,\,x)}{\, f(T,\,x_1) \,} \leq a,  \quad -f_T(T,\,x)>0, \quad
\max_{(T,\, x) \in D} \left\{  -f_T(T,\,x) \right\}  \leq M_T  \big\}.  
\end{eqnarray}
Here,  the norm of the Banach space $C(D)$ is given by
\[
\| g \|=\sup_{(T,\, \xi) \in D} | \, g(T,\, \xi) \,  |, \quad g \in C(D),
\]
and
\[
M_T=\frac{4a \, U_2 \, \displaystyle{ \left( \max_{z \geq 0} \frac{z}{\, \cosh z \,} \right)^2 } }{ \, \varepsilon \, U_1 \, \displaystyle{
\left( \tanh \frac{\varepsilon}{\, 2T_c \,}-\frac{\varepsilon}{\, 2T_c \,} \frac{1}{\cosh^2 \frac{\varepsilon}{\, 2T_c \,} } \right) \int_I \frac{d\xi}{\, \left( \xi^2+\Delta_2(0)^2 \right)^{3/2} \,} \, } } \quad  (>0).
\]
Note that
\[
\sup_{f \in W} \left[ \max_{(T,\, x) \in D} \left\{  -f_T(T,\,x) \right\} \right] = M_T.
\]

\begin{remark}\label{rmk:ffa}
The inequality $f(T,\,x) / f(T,\,x_1) \leq a$ in the definition of $W$ is not defined at $T=T_c$ since $f(T_c,\,x) = 0$. For $T<T_c$, there is a $T_1$ \   ($T<T_1<T_c$) such that $f(T,\,x)=(T-T_c) \, f_T(T_1,\,x)$. Therefore, $f(T_c,\,x) / f(T_c,\,x_1)$ is defined to be $f_T(T_c,\,x) / f_T(T_c,\,x_1)$.
\end{remark}

\begin{remark}
The conditions imposed in the previous papers of the present author \cite[Condition (C)]{watanabe-seven} and \cite[Condition (C)]{watanabe-five} were very complicated, and so it was very tough to show the existence, uniqueness and smoothness of the solution to the BCS-Bogoliubov gap equation \eqref{eqn:bcseq}. Instead, we impose the simple condition that $f \in C^2(D)$ and $f(T_c,\,x) = 0$ in the definition of the subset $W$ (see \eqref{eqn:w}). Thanks to this simple condition, it is straightforward to show the existence, uniqueness and smoothness of the solution.
\end{remark}

Let us remind here that for $T \in [T_0,\, T_c]$ (see \eqref{eqn:tzero}),
\[
\int_I \frac{\, U_1 \,}{\,\xi\,} \, \tanh \frac{\xi}{\,2T \,}\,d\xi <1.
\]
Note that $a \geq 1$ and that $U(x,\,\xi) \geq U_1$ at all $(x,\,\xi) \in I^2$. We then let the potential $U(\cdot,\,\cdot)$ satisfy
\begin{equation}\label{eqn:condition-two}
a^{1/4} \max_{(T,\, x) \in D} \left[ \, 
\int_I \frac{\, U(x,\, \xi) \,}{\xi} \,  \tanh \frac{\xi}{\, 2T\,}\,d\xi \, \right] \leq 1.
\end{equation}

\begin{remark}
The following two theorems hold true not only when the potential $U(\cdot,\,\cdot)$ in the BCS-Bogoliubov gap equation \eqref{eqn:bcseq} is a positive constant, but also when $U(\cdot,\,\cdot)$ is a function. See Remark \ref{rmk:constantpotential} below. 
\end{remark}

We denote by $\overline{W}$ the closure of $W$ with respect to the norm $\| \cdot \|$ mentioned above. The following are our main results.

\begin{theorem}\label{thm:main}
Let the potential $U(\cdot,\,\cdot)$ in the BCS-Bogoliubov gap equation \eqref{eqn:bcseq} satisfy \eqref{eqn:potential} and \eqref{eqn:condition-two}. Let $W$ be as in \eqref{eqn:w}. Then there is a unique fixed point $f_0 \in \overline{W}$ of our operator $A: \, \overline{W} \to \overline{W}$. Therefore, there is a unique nonnegative solution $u_0=\sqrt{f_0}$ to the BCS-Bogoliubov gap equation \eqref{eqn:bcseq}.
\end{theorem}

Let $f_0$ be the fixed point given by Theorem \ref{thm:main} in the following two remarks, where several properties of the solution $u_0=\sqrt{f_0}$ to the BCS-Bogoliubov gap equation \eqref{eqn:bcseq} are pointed out. Suppose $f_0 \in W$. If $f_0$ is an accumulating point of $W$, then $f_0$ can be approximated by an element $f \in W$, and the very element $\sqrt{f}$ satisfies the following properties instead of $u_0$. 

\begin{remark}
\   $u_0 \in C^2([T_0,\, T_1] \times I)$, where $T_1>0$ is arbitrary as long as $T_1<T_c$. Since $f_0(T_c,\, x)=0$ and $(\partial f_0/\partial T)(T,\, x)<0$ at $T$ in a neighborhood of $T_c$, it follows that $u_0(T_c,\, x)=0$ and $(\partial u_0/\partial T)(T,\, x)<0$ at $T$ in a neighborhood of $T_c$. Moreover, $(\partial u_0/\partial T)(T,\, x)\to -\infty$ as $T \uparrow T_c$. But $(\partial u_0^2/\partial T)(T,\, x)\to (\partial f_0/\partial T)(T_c,\, x)$ as $T \uparrow T_c$.
\end{remark}

\begin{remark}
The inequalities  $\Delta_1(T) \leq u_0(T,\,x) \leq \Delta_2(T)$ and $\displaystyle{ \frac{\, u_0(T,\, x)\, }{\, u_0(T,\, x_1) \,} \leq \sqrt{a}}$ hold.
\end{remark}

\medskip

In order to show that the transition from a normal conducting state to a superconducting state at $T=T_c$ is of the second-order, we need to deal with the thermodynamic potential $\Omega$ and differentiate it with respect to the temperature $T$ twice. Note that the thermodynamic potential $\Omega$ has the fixed point $f_0 \in \overline{W}$ given by Theorem \ref{thm:main} in its form, not the solution $\sqrt{ f_0 }$ to the BCS-Bogoliubov gap equation \eqref{eqn:bcseq} in its form. As mentioned before, this is why we treat the squared BCS-Bogoliubov gap equation, not the equation itself . See \cite[(1.5), (1.6)]{watanabe-seven} and \cite[(1.6)]{watanabe-five} for the form of the thermodynamic potential $\Omega$. See also \cite[Definition 1.10]{watanabe-five} for the operator-theoretical definition of the second-order phase transition. 

\begin{theorem}\label{thm:maintwo}
Let the potential $U(\cdot,\,\cdot)$ in the BCS-Bogoliubov gap equation \eqref{eqn:bcseq} satisfy \eqref{eqn:potential} and \eqref{eqn:condition-two}. Let $W$ be as in \eqref{eqn:w}. Then the transition from a normal conducting state to a superconducting state at $T=T_c$ is of the second-order.
\end{theorem}

%%%%%%%%%%%%%%%%%%%%%%%%%%%%%%%%%%%%%%%%%%%%%%%%%%%%%%%%%%%%%%%%% 3
\section{Proofs of Theorems \ref{thm:main} and \ref{thm:maintwo}}

\begin{lemma}
\rm{(1)} \    The subset $W$ is a bounded and convex subset of the Banach space $C(D)$. \\ 
\rm{(2)} \    The closure $\overline{W}$ is a bounded, closed and convex subset of the Banach space $C(D)$.
\end{lemma}

\begin{proof}
\rm{(1)} \   Note that the function $T \mapsto \Delta_2(T)^2$ is strictly decreasing (see \cite[Proposition 1.2]{watanabe-one}). Therefore, $W$ is bounded since $f(T,\,x) \leq \Delta_2(T)^2 \leq \Delta_2(0)^2$ for every $f \in W$. In order to show that $W$ is convex, it suffices to show that
\[
\frac{tf(T,\,x)+(1-t)g(T,\,x)}{\, tf(T,\,x_1)+(1-t)g(T,\,x_1)\,} \leq a.
\]
Here, $t \in [0, \, 1]$ and $f, \, g \in W$. Let $T \not= T_c$. Since $f(T,\,x) \leq a f(T,\,x_1)$ and $g(T,\,x) \leq a g(T,\,x_1)$, it follows
\[
\frac{tf(T,\,x)+(1-t)g(T,\,x)}{\, tf(T,\,x_1)+(1-t)g(T,\,x_1)\,} \leq
\frac{t \, af(T,\,x_1)+(1-t)\, ag(T,\,x_1)}{\, tf(T,\,x_1)+(1-t)g(T,\,x_1)\,}=a.
\]
Next let $T=T_c$. We remind Remark \ref{rmk:ffa} here. Then
\begin{eqnarray*}
& & \frac{tf(T_c,\,x)+(1-t)g(T_c,\,x)}{\, tf(T_c,\,x_1)+(1-t)g(T_c,\,x_1)\,}=
\frac{tf_T(T_c,\,x)+(1-t)g_T(T_c,\,x)}{\, tf_T(T_c,\,x_1)+(1-t)g_T(T_c,\,x_1)\,} \\
&\leq& \frac{t \, af_T(T_c,\,x_1)+(1-t)\, ag_T(T_c,\,x_1)}{\, tf_T(T_c,\,x_1)+(1-t)g_T(T_c,\,x_1)\,}=a.
\end{eqnarray*}
Therefore, $W$ is convex. \\
\rm{(2)} We have only to show that $\overline{W}$ is convex. Let $f, \, g \in \overline{W}$. Then there are $\{ f_n \}, \, \{ g_n \} \subset W$ satisfying $f_n \to f$ and $g_n \to g$ in the Banach space $C(D)$. Since $W$ is convex, $tf_n+(1-t)g_n \in W$ for $t \in [0, \, 1]$.
\[
\left\| \, \{ tf+(1-t)g \}- \{ tf_n+(1-t)g_n \} \, \right\| \leq
t\left\| \, f-f_n \, \right\|+(1-t)\left\| \, g-g_n \, \right\| \to 0
\]
as $n \to \infty$. Thus $tf+(1-t)g \in \overline{W}$, and hence $\overline{W}$ is convex.
\end{proof}

We next show that $A: W \to W$.

\begin{lemma}\label{lem:equicon}
Let $f \in W$. Then $Af$ is continuous on $D$.
\end{lemma}

\begin{proof}
Let $(T,\, x), \, (T_1,\, x_1) \in D$, and suppose $T<T_1<T_c$. We can deal with the case where $T_1=T_c$ similarly . Then
\[
| Af(T,\, x)-Af(T_1,\, x_1) | \leq | Af(T,\, x)-Af(T_1,\, x) |+| Af(T_1,\, x)-Af(T_1,\, x_1) |.
\]
\textit{Step 1}. \   A straightforward calculation gives
\[
Af(T,\, x)-Af(T_1,\, x)=\int_I U(x,\,\eta) \, I_1 \, d\eta \, \int_I U(x,\,\xi)
\left\{ I_2+I_3+I_4  \right\} \, d\xi,
\]
where
\begin{eqnarray*}
I_1 &=& \sqrt{ \, \frac{f(T,\, \eta)}{\, \eta^2+f(T,\, \eta)\,} \,}
 \tanh \frac{\,\sqrt{\,\eta^2+f(T,\, \eta)\,}\,}{2T} +
\sqrt{ \, \frac{f(T_1,\, \eta)}{\, \eta^2+f(T_1,\, \eta)\,} \,} 
 \tanh \frac{\,\sqrt{\,\eta^2+f(T_1,\, \eta)\,}\,}{2T_1}, \\
I_2 &=& \frac{f(T,\, \xi)-f(T_1,\, \xi)}{ \, \sqrt{ f(T,\, \xi) }+\sqrt{ f(T_1,\, \xi) } \, }
 \frac{1}{\, \sqrt{ \xi^2+f(T,\, \xi) \,} \,} \tanh \frac{\,\sqrt{\,\xi^2+f(T,\, \xi)\,}\,}{2T},\\
I_3 &=& \sqrt{ f(T_1,\, \xi) } \left\{ \frac{1}{\, \sqrt{ \xi^2+f(T,\, \xi) \,} \,} \tanh \frac{\,\sqrt{\,\xi^2+f(T,\, \xi)\,}\,}{2T} \right. \\
& & \left. \qquad \qquad \qquad \qquad -\frac{1}{\, \sqrt{ \xi^2+f(T_1,\, \xi) \,} \,}
 \tanh \frac{\,\sqrt{\,\xi^2+f(T_1,\, \xi)\,}\,}{2T} \right\},\\
I_4 &=& \sqrt{ \, \frac{f(T_1,\, \xi)}{\, \xi^2+f(T_1,\, \xi)\,} \,} 
 \left\{ \tanh \frac{\,\sqrt{\, \xi^2+f(T_1,\, \xi)\,}\,}{2T} -
 \tanh \frac{\,\sqrt{\, \xi^2+f(T_1,\, \xi)\,}\,}{2T_1} \right\}.
\end{eqnarray*}
The function $f$ is continuous since $f \in W$. Therefore, for an arbitrary $\varepsilon_1>0$, there is a $\delta>0$ such that if $|T_2-T_3|+|x_2-x_3|<\delta$, then $| f(T_2,\, x_2)-f(T_3,\, x_3) |<\varepsilon_1$. Here, $(T_2,\, x_2), (T_3,\, x_3) \in D$ are arbitrary and the $\delta>0$ does not depend on $(T_2,\, x_2), (T_3,\, x_3)$ since $f$ is uniformly continuous on $D$. Since $f(T,\, \eta)/ f(T,\, \xi) \leq a$ by \eqref{eqn:w},
\begin{eqnarray*}
& & \left| \int_I U(x,\,\eta) \, I_1 \, d\eta \times \int_I U(x,\,\xi) \, I_2 \, d\xi \right| \\
&\leq& 2 \, \frac{ \, U_2^2 \,}{U_1^2} \sqrt{a} \left\{ \int_I 
 \frac{U_1}{\, \sqrt{ \eta^2+\Delta_1(T)^2 \,} \,} 
 \tanh \frac{\,\sqrt{\,\eta^2+\Delta_1(T)^2 \,}\,}{2T} \,d\eta \right\}^2 
 \left| f(T,\, \xi_1)-f(T_1,\, \xi_1) \right| \\
&\leq& 2 \, \frac{ \, U_2^2 \,}{U_1^2} \sqrt{a} \, \varepsilon_1,
\end{eqnarray*}
where $|T-T_1|<\delta$ with some $\xi_1 \in I$. Note that (see \eqref{eqn:delta1})
\[
\int_I \frac{U_1}{\, \sqrt{ \eta^2+\Delta_1(T)^2 \,} \,} 
 \tanh \frac{\,\sqrt{\,\eta^2+\Delta_1(T)^2 \,}\,}{2T} \,d\eta = 1, \quad T \in [0,\, \tau_1]
\]
and that
\[
\int_I \frac{U_1}{\, \sqrt{ \eta^2+0^2 \,} \,} 
 \tanh \frac{\,\sqrt{\,\eta^2+0^2 \,}\,}{2T} \,d\eta < 1, \quad T \in (\tau_1,\, T_c]
\]
with $\Delta_1(T)=0$ at $T \in [\tau_1,\, T_c]$. Since $f(T,\, \xi)>f(T_1,\, \xi)$ \   $(T<T_1)$,
\begin{eqnarray*}
& & \left| \int_I U(x,\,\eta) \, I_1 \, d\eta \times \int_I U(x,\,\xi) \, I_3 \, d\xi \right| \\
&\leq& U_2 \, \Delta_2(0) \left\{ \int_I \frac{U_2}{\, \sqrt{ \eta^2+\Delta_2(T)^2 \,} \,} 
 \tanh \frac{\,\sqrt{\,\eta^2+\Delta_2(T)^2 \,}\,}{2T} \,d\eta \right. \\
& & \qquad \left. +\int_I \frac{U_2}{\, \sqrt{ \eta^2+\Delta_2(T_1)^2 \,} \,} 
 \tanh \frac{\,\sqrt{\,\eta^2+\Delta_2(T_1)^2 \,}\,}{2T_1} \,d\eta \right\} \\
& & \qquad \times \int_I \frac{1}{\, \xi^2 \,} \,d\xi \, 
\left| f(T,\, \xi_1)-f(T_1,\, \xi_1) \right|  \\
&\leq& 2 \, \frac{ \, U_2 \, \Delta_2(0)\,}{\varepsilon} \, \varepsilon_1,
\end{eqnarray*}
where $|T-T_1|<\delta$ with some $\xi_1 \in I$. Similarly,
\begin{eqnarray*}
& & \left| \int_I U(x,\,\eta) \, I_1 \, d\eta \times \int_I U(x,\,\xi) \, I_4 \, d\xi \right| \\
&\leq& 2 U_2 \, \Delta_2(0) \, \left( \max_{z \geq 0} \frac{z}{\, \cosh z \,} \right)^2
 \left\{ \int_I \frac{U_2}{\, \sqrt{ \eta^2+\Delta_2(T)^2 \,} \,} 
 \tanh \frac{\,\sqrt{\,\eta^2+\Delta_2(T)^2 \,}\,}{2T} \,d\eta \right. \\
& & \qquad \left. +\int_I \frac{U_2}{\, \sqrt{ \eta^2+\Delta_2(T_1)^2 \,} \,} 
 \tanh \frac{\,\sqrt{\,\eta^2+\Delta_2(T_1)^2 \,}\,}{2T_1} \,d\eta \right\} \\
& & \qquad \times \int_I \frac{1}{\, \xi \,} \,d\xi \, |T-T_1| \\
&\leq& 4 U_2 \, \Delta_2(0) \, \left( \max_{z \geq 0} \frac{z}{\, \cosh z \,} \right)^2 \, (\ln \varepsilon) \, |T-T_1|<\varepsilon_1,
\end{eqnarray*}
where
\[
|T-T_1|<\delta_1=\frac{\varepsilon_1}{\, 4 \, U_2 \, \Delta_2(0) 
 \left( \max_{z \geq 0} \frac{z}{\, \cosh z \,} \right)^2 \, \ln \varepsilon \,}.
\]
Thus
\[
| Af(T,\, x)-Af(T_1,\, x) | \leq \left( 2 \, \frac{ \, U_2^2 \,}{U_1^2} \sqrt{a}+
 2 \, \frac{ \, U_2 \, \Delta_2(0)\,}{\varepsilon}+1 \right) \, \varepsilon_1,
\]
where $|T-T_1|<\min (\delta, \, \delta_1)$. \\
\textit{Step 2}. \  By hypothesis, the potential $U(\cdot,\,\cdot)$ is continuous on the compact set $I^2$, and hence $U(\cdot,\,\cdot)$ is uniformly continuous. Therefore, for an arbitrary $\varepsilon_1>0$, there is a $\delta_2>0$ such that if $|x-x_1|<\delta_2$, then $| U(x,\, \eta)-U(x_1,\, \eta) |<\varepsilon_1$. Note that the $\delta_2$ does not depend on $f \in W$. A straightforward calculation gives
\begin{eqnarray*}
& & \left| Af(T_1,\, x)-Af(T_1,\, x_1) \right| \\
&\leq& \int_I \left\{ U(x,\,\eta)+U(x_1,\,\eta) \right\}
 \sqrt{ \, \frac{f(T_1,\, \eta)}{\, \eta^2+f(T_1,\, \eta)\,} \,} 
 \tanh \frac{\,\sqrt{\,\eta^2+f(T_1,\, \eta)\,}\,}{2T_1} \, d\eta \\
& & \quad  \times \int_I \left| U(x,\,\xi)-U(x_1,\,\xi) \right| 
 \sqrt{ \, \frac{f(T_1,\, \xi)}{\, \xi^2+f(T_1,\, \xi)\,} \,} 
 \tanh \frac{\,\sqrt{\, \xi^2+f(T_1,\, \xi)\,}\,}{2T_1} \, d\xi \\
&\leq& 2 \int_I \, U_2 \, \sqrt{ \, \frac{\Delta_2(T_1)^2}{\, \eta^2+\Delta_2(T_1)^2\,} \,} 
 \tanh \frac{\,\sqrt{\,\eta^2+\Delta_2(T_1)^2 \,}\,}{2T_1} \, d\eta \\
& & \quad  \times \int_I \, \left| U(x,\,\xi)-U(x_1,\,\xi) \right|  \, 
 \sqrt{ \, \frac{\Delta_2(T_1)^2}{\, \eta^2+\Delta_2(T_1)^2\,} \,} 
 \tanh \frac{\,\sqrt{\,\eta^2+\Delta_2(T_1)^2 \,}\,}{2T_1} \, d\xi \\
&\leq& 2 \frac{\, \Delta_2(0)^2 \,}{U_2} \, \varepsilon_1,
\end{eqnarray*}
where $|x-x_1|<\delta_2$. \\
\textit{Step 3}. \   Steps 1 and 2 thus imply
\[
| Af(T,\, x)-Af(T_1,\, x_1) | \leq \left( 2 \, \frac{ \, U_2^2 \,}{U_1^2} \sqrt{a}+
 2 \, \frac{ \, U_2 \, \Delta_2(0)\,}{\varepsilon}+1+
2 \frac{\, \Delta_2(0)^2 \,}{U_2} \right) \, \varepsilon_1,
\]
where $|T-T_1|+|x-x_1|<\min (\delta, \, \delta_1, \delta_2)$. Therefore, $Af$ is continuous on $D$.
\end{proof}  

\begin{lemma} \   Let $f \in W$. \\
\rm{(1)} \    $Af$ is partially differentiable with respect to both $T$ and $x$. Its first-order partial derivatives $(Af)_T$ and $(Af)_x$ are both continuous on $D$. Therefore, $Af \in C^1(D)$. \\
\rm{(2)} \    $Af$ is twice partially differentiable with respect to both $T$ and $x$.  Its second-order partial derivatives $(Af)_{TT}$, $(Af)_{Tx}=(Af)_{xT}$ and $(Af)_{xx}$ are all continuous on $D$. Therefore, $Af \in C^2(D)$.
\end{lemma}

\begin{proof}
(1) \   Let us show that $Af$ is partially differentiable with respect to $T$ at $(T_c,\,x_0) \in D$. Note that $Af(T_c,\,x_0)=0$. Let $T<T_c$. It follows from $f(T_c,\, \xi)=0$ (see \eqref{eqn:w}) that
\[
f(T,\, \xi)=f(T_c,\, \xi)+(T-T_c)f_T(T_1,\, \xi)=(T_c-T)\left( -f_T(T_1,\, \xi) \right)
\]
for some $T_1$ \   $(T<T_1<T_c)$. Then
\begin{eqnarray*}
\frac{\, Af(T_c,\,x_0)-Af(T,\,x_0) \,}{T_c-T} &=&
-\left( \int_I U(x_0,\,\xi) \sqrt{ \, \frac{f(T,\, \xi)/(T_c-T)}{\, \xi^2+f(T,\, \xi)\,} \,} \, \tanh \frac{\,\sqrt{\,\xi^2+f(T,\, \xi)\,}\,}{2T}\, d\xi \right)^2 \\
&=& -\left( \int_I U(x_0,\,\xi) \sqrt{ \, \frac{-f_T(T_1,\, \xi)}{\, \xi^2+f(T,\, \xi)\,} \,} \, \tanh \frac{\,\sqrt{\,\xi^2+f(T,\, \xi)\,}\,}{2T}\, d\xi \right)^2.
\end{eqnarray*}
Since $T$ is in a neighborhood of $T_c$, we let $T \geq T_c/2$. Therefore,
\[
\sqrt{ \, \frac{-f_T(T_1,\, \xi)}{\, \xi^2+f(T,\, \xi)\,} \,} \, \tanh \frac{\,\sqrt{\,\xi^2+f(T,\, \xi)\,}\,}{2T} \leq  \frac{ \, \sqrt{ M_T} \, }{\xi} \, \tanh \frac{\, \xi \,}{T_c},
\]
where the right side is independent of $T$ and is Lebesgue integrable on $I$. Thus, as $T \uparrow T_c$,
\[
\frac{\, Af(T_c,\,x_0)-Af(T,\,x_0) \,}{T_c-T} \to
-\left( \int_I U(x_0,\,\xi) \frac{ \, \sqrt{ -f_T(T_c,\, \xi) } \, }{\xi} \, \tanh \frac{\, \xi \,}{\, 2T_c \,} \, d\xi \right)^2.
\]
Therefore, $Af$ is partially differentiable with respect to $T$ at $(T_c,\,x_0)$, and
\begin{equation}\label{eqn:afttc}
(Af)_T(T_c,\,x_0)=-\left( \int_I U(x_0,\,\xi) \frac{ \, \sqrt{ -f_T(T_c,\, \xi) } \, }{\xi} \, \tanh \frac{\, \xi \,}{\, 2T_c \,} \, d\xi \right)^2.
\end{equation}
We next show that $(Af)_T$ is continuous at $(T_c,\,x_0)$. Here,
\begin{eqnarray}\label{eqn:afT}
(Af)_T(T,\,x) &=& \int_I U(x,\,\eta) \sqrt{ \, \frac{f(T,\, \eta)}{\, \eta^2+f(T,\, \eta)\,} \,} \, \tanh \frac{\,\sqrt{\,\eta^2+f(T,\, \eta)\,}\,}{2T}\, d\eta \\ \nonumber
& & \quad \times \int_I U(x,\,\xi) \left( J_1+J_2+J_3 \right) \, d\xi,
\end{eqnarray}
where
\begin{eqnarray*}
J_1 &=& \frac{\, f_T(T,\,\xi) \,}{\, \sqrt{ f(T,\,\xi)} \,} \,
 \frac{\xi^2}{\, \left\{ \, \xi^2+f(T,\, \xi) \, \right\}^{3/2}\, }
 \tanh \frac{ \, \sqrt{ \, \xi^2+f(T,\, \xi) \, } \,}{2T}, \nonumber \\
J_2 &=& \frac{\sqrt{ f(T,\, \xi) } \, f_T(T,\, \xi)}{\, 2T \left\{ \, \xi^2+f(T,\, \xi) \, \right\} 
 \cosh^2 \frac{ \, \sqrt{ \, \xi^2+f(T,\, \xi) \, }  \, }{2T}  \,},  \nonumber \\
J_3 &=& -\frac{ \sqrt{ f(T,\, \xi) } }{\, T^2 \cosh^2 \frac{ \, \sqrt{ \, \xi^2+f(T,\, \xi) \, }  \, }{2T}  \,}. \nonumber
\end{eqnarray*}
Note that
\begin{eqnarray}\label{eqn:AfT}
(Af)_T(T,\,x)-(Af)_T(T_c,\,x_0) &=& (Af)_T(T,\,x)-(Af)_T(T_c,\,x) \\ \nonumber
& & \quad + (Af)_T(T_c,\,x)-(Af)_T(T_c,\,x_0).
\end{eqnarray}
In order to show that $(Af)_T(T,\,x) \to (Af)_T(T_c,\,x)$ as $T \uparrow T_c$, we show that as $T \uparrow T_c$,
\begin{eqnarray*}
\int_I U(x,\,\eta) \sqrt{ \, \frac{f(T,\, \eta)}{\, \eta^2+f(T,\, \eta)\,} \,} \, \tanh \frac{\,\sqrt{\,\eta^2+f(T,\, \eta)\,}\,}{2T}\, d\eta \int_I U(x,\,\xi) \, J_1 \, d\xi
&\to& (Af)_T(T_c,\,x), \\
\int_I U(x,\,\eta) \sqrt{ \, \frac{f(T,\, \eta)}{\, \eta^2+f(T,\, \eta)\,} \,} \, \tanh \frac{\,\sqrt{\,\eta^2+f(T,\, \eta)\,}\,}{2T}\, d\eta \int_I U(x,\,\xi) \, (J_2+J_3) \, d\xi
&\to& 0.
\end{eqnarray*}
A straightforward calculation gives
\[
U(x,\,\eta) \sqrt{ \, \frac{f(T,\, \eta)}{\, \eta^2+f(T,\, \eta)\,} \,} \, \tanh \frac{\,\sqrt{\,\eta^2+f(T,\, \eta)\,}\,}{2T} \; U(x,\,\xi) \, J_1 \leq U_2^2 \, \sqrt{a}
 \left( \frac{1}{\, \eta \,} \tanh \frac{\, \eta \,}{T_c} \right)^2 M_T.
\]
Here, we assumed $T \geq T_c/2$. The right side of this inequality is independent of $T$ and is Lebesgue integrable on $I^2$, and so \   (as $T \uparrow T_c$)
\[
\int_I U(x,\,\eta) \sqrt{ \, \frac{f(T,\, \eta)}{\, \eta^2+f(T,\, \eta)\,} \,} \, \tanh \frac{\,\sqrt{\,\eta^2+f(T,\, \eta)\,}\,}{2T}\, d\eta \int_I U(x,\,\xi) \, J_1 \, d\xi
\to (Af)_T(T_c,\,x).
\]
Similarly we can show that
\[
\int_I U(x,\,\eta) \sqrt{ \, \frac{f(T,\, \eta)}{\, \eta^2+f(T,\, \eta)\,} \,} \, \tanh \frac{\,\sqrt{\,\eta^2+f(T,\, \eta)\,}\,}{2T}\, d\eta \int_I U(x,\,\xi) \, (J_2+J_3) \, d\xi
\to 0
\]
as $T \uparrow T_c$. Moreover, we have similarly that
\[
(Af)_T(T_c,\,x) \to (Af)_T(T_c,\,x_0)
\]
as $x \to x_0$. It thus follows from \eqref{eqn:AfT} that $(Af)_T$ is continuous at $(T_c,\,x_0)$. Similarly we can show the rest of (1), and (2). Note that $(Af)_{TT}$ is given as follows.
\begin{eqnarray*}
\quad (Af)_{TT}(T,\,x) &=& \frac{1}{\, 2 \,} \left\{ \int_I U(x,\,\eta) \, \left( J_1+J_2+J_3 \right)
 \, d\eta \right\}^2 \\
& & +\int_I U(x,\,\eta) \sqrt{ \, \frac{f(T,\, \eta)}{\, \eta^2+f(T,\, \eta)\,} \,} \, \tanh \frac{\,\sqrt{\,\eta^2+f(T,\, \eta)\,}\,}{2T}\, d\eta \nonumber \\
& & \times \int_I U(x,\,\xi) \left\{ K_1+
 \frac{1}{\, \cosh^2 \frac{ \, \sqrt{ \, \xi^2+f(T,\, \xi) \, }  \, }{2T}  \,}
\left( K_2+K_3+K_4+K_5 \right) \right\} \, d\xi, \nonumber
\end{eqnarray*}
where
\begin{eqnarray*}
K_1 &=& \left\{ \frac{\, f_{TT}(T,\,\xi) \,}{\, \sqrt{ f(T,\,\xi)} \,}
 -\frac{\, f_T(T,\,\xi)^2 \,}{\, 2\sqrt{ f(T,\,\xi)}^3 \,}
 -\frac{\, 3 f_T(T,\,\xi)^2 \,}{\, 2\sqrt{ f(T,\,\xi)} \left( \, \xi^2+f(T,\, \xi) \, \right) \,}
 \right\} \\
& & \qquad \times \frac{\xi^2}{\, \left\{ \, \xi^2+f(T,\, \xi) \, \right\}^{3/2}\, }
 \tanh \frac{ \, \sqrt{ \, \xi^2+f(T,\, \xi) \, } \,}{2T}, \\
K_2 &=& \frac{\, f_T(T,\,\xi) \,}{\, 2\sqrt{ f(T,\,\xi)} \,}
 \frac{\xi^2}{\, \xi^2+f(T,\, \xi) \, }
 \left\{ \frac{\, f_T(T,\,\xi) \,}{\, 2T \, ( \xi^2+f(T,\, \xi) ) \,}
  -\frac{1}{\, T^2 \,} \right\}, \\
K_3 &=& \frac{\, f_T(T,\,\xi) \,}{\, 2\sqrt{ f(T,\,\xi)} \,}
  \left\{ \frac{\, f_T(T,\,\xi) \,}{\, 2T \, ( \xi^2+f(T,\, \xi) ) \,}
  -\frac{1}{\, T^2 \,} \right\}, \\
K_4 &=& \sqrt{ f(T,\,\xi) \,} \left\{ 
 \frac{\, f_{TT}(T,\,\xi) \,}{\, 2T \, ( \xi^2+f(T,\, \xi) ) \,}
 -\frac{\, f_T(T,\,\xi) \,}{\, 2T^2 \, ( \xi^2+f(T,\, \xi) ) \,}
 -\frac{\, (f_T(T,\,\xi))^2 \,}{\, 2T \, ( \xi^2+f(T,\, \xi) )^2 \,}
 +\frac{2}{\, T^3 \,} \right\}, \\
K_5 &=& -\sqrt{ f(T,\,\xi) \,} \, \sqrt{ \, \xi^2+f(T,\, \xi) \, }
 \left\{ \frac{\, f_T(T,\,\xi) \,}{\, 2T \, ( \xi^2+f(T,\, \xi) ) \,}
  -\frac{1}{\, T^2 \,} \right\}^2 \tanh \frac{ \, \sqrt{ \, \xi^2+f(T,\, \xi) \, } \,}{2T}.
\end{eqnarray*}
\end{proof} 

A proof similar to that of \cite[Lemma 3.4]{watanabe-one} gives the following.
\begin{lemma}\label{lm:delta}
Let $f \in W$. Then $\Delta_1(T)^2 \leq Af(T,\,x) \leq \Delta_2(T)^2$ at each $(T,\,x) \in D$.
\end{lemma}

\begin{lemma}
Let $f \in W$. Then $Af(T_c,\, x)=0$ \   $(x \in I)$, and
\[
\frac{Af(T,\,x)}{\,Af(T,\,x_1) \,} \leq a.
\]
\end{lemma}

\begin{proof} \  By \eqref{eqn:w},
\[
Af(T_c,\,x)=\left( \int_I U(x,\,\xi) \sqrt{ \, \frac{f(T_c,\, \xi)}{\, \xi^2+f(T_c,\, \xi)\,} \,} \, \tanh \frac{\,\sqrt{\,\xi^2+f(T_c,\, \xi)\,}\,}{2T_c}\, d\xi \right)^2=0.
\]
Next, it follows from \eqref{eqn:ourop} that at $T \in [0,\, T_c)$,
\[
\frac{Af(T,\,x)}{\,Af(T,\,x_1) \,} = \left( \frac{U(x,\, \xi_1)}{\, U(x_1,\, \xi_2) \,} \right)^2 \leq \left( \frac{ \displaystyle{ \max_{(x,\,\xi) \in I^2} U(x,\,\xi) }}{\, \displaystyle{ \min_{(x,\,\xi) \in I^2} U(x,\,\xi) } \,} \right)^2 = a,
\]
where $\xi_1, \, \xi_2 \in I$. It then follows from Remark \ref{rmk:ffa} and \eqref{eqn:afttc} that at $T=T_c$,
\[
\frac{Af(T_c,\, x)}{\,Af(T_c,\, x_1) \,}= \frac{(Af)_T(T_c,\, x)}{\, (Af)_T(T_c,\, x_1) \,} \leq \left( \frac{ \displaystyle{ \max_{(x,\,\xi) \in I^2} U(x,\,\xi) }}{\, \displaystyle{ \min_{(x,\,\xi) \in I^2} U(x,\,\xi) } \,} \right)^2 =a.
\]
The result follows.
\end{proof}

\begin{lemma} \   For $f \in W$, \quad $\displaystyle{ -(Af)_T(T,\,x)>0. }$
\end{lemma}

\begin{proof}
It follows immediately from \eqref{eqn:afT} that $-(Af)_T(T,\,x)>0$.\end{proof}

\begin{lemma} \  For $f \in W$,
\[
\sup_{f \in W} \left[ \max_{(T,\, x) \in D} \left\{  -(Af)_T(T,\,x) \right\} \right]
\leq \sup_{f \in W} \left[ \max_{(T,\, x) \in D} \left\{  -f_T(T,\,x) \right\} \right] \; \big(=M_T \big).
\]
\end{lemma}

\begin{proof}
From \eqref{eqn:afT} it follows that
\[
-(Af)_T(T,\,x) \leq \sqrt{a}\, A \left\{ B\sup_{f \in W} \left[ \max_{(T,\, x) \in D} \left\{  -f_T(T,\,x) \right\} \right]+C \right\},
\]
where
\begin{eqnarray*}
A &=& \int_I \frac{ \, U(x,\,\eta) \,}{\, \sqrt{ \, \eta^2+f(T,\, \eta)\,} \,} \, \tanh \frac{\,\sqrt{\,\eta^2+f(T,\, \eta)\,}\,}{2T}\, d\eta, \\
B &=& \int_I \frac{ \, U(x,\,\xi) \,}{\, \sqrt{ \, \xi^2+f(T,\, \xi)\,} \,} \left\{ 
\frac{\xi^2}{\, \xi^2+f(T,\, \xi) \, }\,
\tanh \frac{\,\sqrt{\,\xi^2+f(T,\, \xi)\,}\,}{2T}\right. \\
& & \qquad \left. +\frac{f(T,\, \xi)}{\, \xi^2+f(T,\, \xi) \, }\,
\frac{\sqrt{ \, \xi^2+f(T,\, \xi)\,}}{\, 2T \,\cosh^2 \frac{\,\sqrt{\,\xi^2+f(T,\, \xi)\,}\,}{2T} \,}  \right\}\, d\xi, \\
C &=& \int_I U(x,\,\xi) \,
\frac{ \, f(T,\, \xi) \,}{\, T^2 \, \cosh^2 \frac{\,\sqrt{\,\xi^2+f(T,\, \xi)\,}\,}{2T}\, }\, d\xi.
\end{eqnarray*}
Note that the function $\displaystyle{ z \mapsto \frac{\,\tanh z \,}{z} }$ is strictly decreasing at $z>0$. It then follows from \eqref{eqn:condition-two} that
\begin{equation}\label{eqn:aaone}
a^{1/4}A \leq a^{1/4} \max_{(T,\, x) \in D} \left[ \, 
\int_I \frac{\, U(x,\, \eta) \,}{\eta} \,  \tanh \frac{\eta}{\, 2T\,}\,d\eta \, \right] \leq 1.
\end{equation}

First let $T<T_c$. Then $B<A$ since $\displaystyle{ \tanh z>\frac{z}{\cosh z} }$ at $z>0$, $f(T,\, \xi)>0$ and $\xi \geq \varepsilon$. Therefore, $\sqrt{a}\, AB<( a^{1/4}A)^2\leq 1$ by \eqref{eqn:aaone}.  Thus
\[
-(Af)_T(T,\,x) \leq \sqrt{a}\, A \left\{ B\sup_{f \in W} \left[ \max_{(T,\, x) \in D} \left\{  -f_T(T,\,x) \right\} \right]+C \right\} \leq \sup_{f \in W} \left[ \max_{(T,\, x) \in D} \left\{  -f_T(T,\,x) \right\} \right]
\]
as long as
\[
\sup_{f \in W} \left[ \max_{(T,\, x) \in D} \left\{  -f_T(T,\,x) \right\} \right] \geq
\frac{\sqrt{a} \, AC}{\, 1-\sqrt{a}\, AB \,}.
\]
This inequality holds true since (see the definition of $M_T$ in \eqref{eqn:w})
\[
\frac{\sqrt{a} \, AC}{\, 1-\sqrt{a}\, AB \,} = \frac{\sqrt{a} \, AC}{\, 1-\sqrt{a}\, A(A-B') \,} \leq \frac{\sqrt{a} \, AC}{\, \sqrt{a}\, AB' \,}=\frac{C}{\,B' \,} \leq M_T
= \sup_{f \in W} \left[ \max_{(T,\, x) \in D} \left\{  -f_T(T,\,x) \right\} \right].
\]
Here,  
\[
B'=\int_I \frac{ \, U(x,\,\xi) \,}{\, \sqrt{ \, \xi^2+f(T,\, \xi)\,} \,} 
\frac{f(T,\, \xi)}{\, \xi^2+f(T,\, \xi) \, }\left\{
\tanh \frac{\,\sqrt{\,\xi^2+f(T,\, \xi)\,}\,}{2T}-\frac{\sqrt{ \, \xi^2+f(T,\, \xi)\,}}{\, 2T \,\cosh^2 \frac{\,\sqrt{\,\xi^2+f(T,\, \xi)\,}\,}{2T} \,}  \right\}\, d\xi.
\]
Note also that $1-\sqrt{a}\, A^2 \geq 0$. Thus, at $T<T_c$,
\[
-(Af)_T(T,\,x) \leq \sup_{f \in W} \left[ \max_{(T,\, x) \in D} \left\{  -f_T(T,\,x) \right\} \right].
\]

Next let $T=T_c$. Then
\begin{eqnarray*}
& &-(Af)_T(T_c,\,x)=\left( \int_I U(x,\,\xi)\frac{\, \sqrt{ -f_T(T_c,\,\xi)}  \,}{\, \xi \,} \, 
\tanh \frac{\, \xi \,}{2T_c}\, d\xi \right)^2 \\
&\leq& \sqrt{a}\left( \int_I U(x,\,\xi)\frac{\, \sqrt{ -f_T(T_c,\,\xi)}  \,}{\, \xi \,} \, 
\tanh \frac{\, \xi \,}{2T_c}\, d\xi \right)^2 \\
&\leq& \sqrt{a}\left( \int_I U(x,\,\xi)\frac{\, 1 \,}{\, \xi \,} \, 
\tanh \frac{\, \xi \,}{2T_c}\, d\xi \right)^2 
\sup_{f \in W} \left[ \max_{(T,\, x) \in D} \left\{  -f_T(T,\,x) \right\} \right] \\
&\leq& \sup_{f \in W} \left[ \max_{(T,\, x) \in D} \left\{  -f_T(T,\,x) \right\} \right].
\end{eqnarray*}
This is because at $T=T_c$,
\[
 \sqrt{a}\, A^2=\left( a^{1/4}\, \int_I \frac{\, U(x,\,\eta) \,}{\, \eta \,} \, 
\tanh \frac{\, \eta \,}{2T_c}\, d\eta \right)^2 \leq 1
\]
by \eqref{eqn:aaone}. Thus
\[
\sup_{f \in W} \left[ \max_{(T,\, x) \in D} \left\{  -(Af)_T(T,\,x) \right\} \right] \leq
\sup_{f \in W} \left[ \max_{(T,\, x) \in D} \left\{  -f_T(T,\,x) \right\} \right] =M_T.
\]
\end{proof}

\begin{remark}\label{rmk:constantpotential}
Let $U(x,\,\xi)=U_1=U_2$ at all $(x,\, \xi) \in I^2$. Then, $a=1$ and $f(T,\, x)=\Delta_1(T)^2=\Delta_2(T)^2$. Moreover, $T_c=\tau_1=\tau_2$ and
\[
a^{1/4}A=\int_I \frac{U_2}{\,\sqrt{\,\xi^2+\Delta_2(T)^2\,}\,} \,
 \tanh \frac{\, \sqrt{\,\xi^2+\Delta_2(T)^2\,}\,}{2T}\,d\xi=1
\]
at all $(T,\, x) \in [0,\, T_c] \times I$ (see \eqref{eqn:delta2}). Therefore, the preceding lemma holds true not only when the potential $U(\cdot,\,\cdot)$ in the BCS-Bogoliubov gap equation \eqref{eqn:bcseq} is a positive constant, but also when $U(\cdot,\,\cdot)$ is a function.
\end{remark}

\begin{lemma}
The set $AW$ is equicontinuous.
\end{lemma}

\begin{proof}
Let $f \in W$. Let $(T,\, x), \, (T_1,\, x_1) \in D$ and suppose $T<T_1<T_c$. We can deal with the case where $T_1=T_c$ similarly. Then
\[ 
| Af(T,\, x)-Af(T_1,\, x_1) | \leq | Af(T,\, x)-Af(T_1,\, x) |+| Af(T_1,\, x)-Af(T_1,\, x_1) |.
\]
The preceding lemma gives
\[
| f(T,\, \xi)-f(T_1,\, \xi) |  = \left| f_T(T_2, \, \xi) \right| \cdot \left| T-T_1 \right| \leq
M_T | T-T_1 |.
\]
Here, $T<T_2<T_1$ and $\xi \in I$. Therefore, a proof similar to that of Lemm \ref{lem:equicon} gives
\begin{eqnarray*}
| Af(T,\, x)-Af(T_1,\, x_1) |  &\leq& \left\{ \left( 2 \, \frac{ \, U_2^2 \,}{U_1^2} \sqrt{a}+ 2 \, \frac{ \, U_2 \, \Delta_2(0)\,}{\varepsilon} \right) M_T \right. \\
& & + \; 4 \, U_2 \, \Delta_2(0) \left( \max_{z \geq 0} \frac{z}{\, \cosh z \,} \right)^2 \, \ln \varepsilon \\
& & \left.+2 \; \frac{\, \Delta_2(0)^2 \,}{U_2}\, 
 \max_{(x,\, \xi) \in I^2} \left\{ U_x(x,\,\xi) \right\}\right\}\, 
\left(  | T-T_1 | +| x-x_1 |  \right),
\end{eqnarray*}
from which the result follows.
\end{proof}

Since $Af(T,\, x)\leq \Delta_2(T)^2 \leq \Delta_2(0)^2$ for $f \in W$ (see Lemma \ref{lm:delta}), the set $AW$ is uniformly bounded. Moreover, $AW$ is equicontinuous by the preceding lemma. We thus have the following.
\begin{lemma}\label{lm:setaw}
\   $\displaystyle{ A: \, W \to W}$, and the set $AW$ is relatively compact.
\end{lemma}

\begin{lemma}\label{lm:acontinuous}
The operator $A:\,  W \to W$ is continuous.
\end{lemma}

\begin{proof}
Let $T<T_c$. Then, for $f, \, g \in W$,
\[
Af(T,\, x)-Ag(T,\, x)=\int_I U(x,\,\eta) \, L_1 \, d\eta \, \int_I U(x,\,\xi)
\left\{ L_2+L_3 \right\} \, d\xi,
\]
where
\begin{eqnarray*}
L_1 &=& \sqrt{ \, \frac{f(T,\, \eta)}{\, \eta^2+f(T,\, \eta)\,} \,}
 \tanh \frac{\,\sqrt{\,\eta^2+f(T,\, \eta)\,}\,}{2T} +
\sqrt{ \, \frac{g(T,\, \eta)}{\, \eta^2+g(T,\, \eta)\,} \,} 
 \tanh \frac{\,\sqrt{\,\eta^2+g(T,\, \eta)\,}\,}{2T}, \\
L_2 &=& \frac{f(T,\, \xi)-g(T,\, \xi)}{ \, \sqrt{ f(T,\, \xi) }+\sqrt{ g(T,\, \xi) } \, }
 \frac{1}{\, \sqrt{ \xi^2+f(T,\, \xi) \,} \,} \tanh \frac{\,\sqrt{\,\xi^2+f(T,\, \xi)\,}\,}{2T},\\
L_3 &=& \sqrt{ g(T,\, \xi) } \left\{ \frac{1}{\, \sqrt{ \xi^2+f(T,\, \xi) \,} \,} \tanh \frac{\,\sqrt{\,\xi^2+f(T,\, \xi)\,}\,}{2T} \right. \\
& & \left. \qquad \qquad \qquad \qquad -\frac{1}{\, \sqrt{ \xi^2+g(T,\, \xi) \,} \,}
 \tanh \frac{\,\sqrt{\,\xi^2+g(T,\, \xi)\,}\,}{2T} \right\}.
\end{eqnarray*}
Since $f(T,\, \eta)/ f(T,\, \xi) \leq a$ and $g(T,\, \eta)/ g(T,\, \xi)\leq a$ by \eqref{eqn:w}, it follows
\begin{eqnarray*}
& &\left| \int_I U(x,\,\eta) \, L_1 \, d\eta \times \int_I U(x,\,\xi) \, L_2 \, d\xi \right| \\
&\leq& 2 \, \frac{ \, U_2^2 \,}{U_1^2} \sqrt{a} \left\{ \int_I 
 \frac{U_1}{\, \sqrt{ \eta^2+\Delta_1(T)^2 \,} \,} 
 \tanh \frac{\,\sqrt{\,\eta^2+\Delta_1(T)^2 \,}\,}{2T} \,d\eta \right\}^2 \| f-g \| \\
&\leq& 2 \, \frac{ \, U_2^2 \,}{U_1^2} \sqrt{a} \, \| f-g \|.
\end{eqnarray*}
Moreover,
\begin{eqnarray*}
& &\left| \int_I U(x,\,\eta) \, L_1 \, d\eta \times \int_I U(x,\,\xi) \, L_3 \, d\xi \right| \\
&\leq& 2 \, \frac{ \, U_2^2 \, \Delta_2(0)^2 \,}{U_1} \int_I 
 \frac{U_1}{\, \sqrt{ \eta^2+\Delta_1(T)^2 \,} \,} 
 \tanh \frac{\,\sqrt{\,\eta^2+\Delta_1(T)^2 \,}\,}{2T} \,d\eta \,
 \int_I \frac{1}{\, \xi^3 \,} \,d\xi \, \| f-g \|  \\
&\leq& \frac{ \, U_2^2 \, \Delta_2(0)^2 \,}{U_1 \, \varepsilon^2} \, \| f-g \|.
\end{eqnarray*}
Therefore, at $T<T_c$,
\[
| Af(T,\, x)-Ag(T,\, x) | \leq \left( 2 \, \frac{ \, U_2^2 \,}{U_1^2} \sqrt{a}+
 \frac{ \, U_2^2 \, \Delta_2(0)^2 \,}{U_1 \, \varepsilon^2} \right) \, \| f-g \|.
\]
Since $Af(T_c,\, x)=Ag(T_c,\, x)=0$, this inequality holds true also at $T=T_c$. Thus
\[
\| Af-Ag \| \leq \left( 2 \, \frac{ \, U_2^2 \,}{U_1^2} \sqrt{a}+
 \frac{ \, U_2^2 \, \Delta_2(0)^2 \,}{U_1 \, \varepsilon^2} \right) \, \| f-g \|.
\]
The result follows.
\end{proof}

We next extend the domain $W$ of our operator $A$ to its closure $\overline{W}$ with respect to the norm $\| \cdot \|$ of the Banach space $C(D)$.

\begin{lemma}\label{lm:overlineW} \quad $\displaystyle{ A: \, \overline{W} \to \overline{W}}$.
\end{lemma}

\begin{proof}
For $f \in \overline{W}$, there is a sequence $\{ f_n \}_{n=1}^{\infty} \subset W$ satisfying $\| f-f_n \| \to 0$ as $n \to \infty$. By the preceding lemma,
\[
\| Af_n-Af_m \| \leq \left( 2 \, \frac{ \, U_2^2 \,}{U_1^2} \sqrt{a}+
 \frac{ \, U_2^2 \, \Delta_2(0)^2 \,}{U_1 \, \varepsilon^2} \right) \, \| f_n-f_m \|.
\]
Therefore, the sequence $\{ Af_n \}_{n=1}^{\infty} \subset W$ is a Cauchy sequence. Hence there is an element $F \in \overline{W}$  satisfying $\| F-Af_n \| \to 0$ as $n \to \infty$. Note that the element $F$ does not depend on how to choose the sequence $\{ f_n \}_{n=1}^{\infty} \subset W$, as shown below. Suppose that there is another sequence $\{ g_n \}_{n=1}^{\infty} \subset W$ satisfying $\| f-g_n \| \to 0$ as $n \to \infty$. Similarly,  the sequence $\{ Ag_n \}_{n=1}^{\infty} \subset W$ becomes a Cauchy sequence, and hence there is an element $G \in \overline{W}$  satisfying $\| G-Ag_n \| \to 0$ as $n \to \infty$. Then
\[
\| F-G \| \leq \| F-Af_n \|+\| Af_n-Ag_n \|+\| Ag_n-G \| \to 0
\]
as $n \to \infty$. Therefore, $F=G$, and hence $F$ does not depend on how to choose the sequence in $W$. Thus we define $F=Af$. The result thus follows.
\end{proof}

\begin{lemma}
For $f \in \overline{W}$,
\[
Af(T,\,x)=\left( \int_I U(x,\,\xi) \sqrt{ \, \frac{f(T,\, \xi)}{\, \xi^2+f(T,\, \xi)\,} \,} \, \tanh \frac{\,\sqrt{\,\xi^2+f(T,\, \xi)\,}\,}{2T}\, d\xi \right)^2.
\]
\end{lemma}

\begin{proof} \  For $f \in \overline{W}$, there is a sequence $\{ f_n \}_{n=1}^{\infty} \subset W$ satisfying $\| f-f_n \| \to 0$ as $n \to \infty$. Since $f$ is Lebesgue integrable on $I$, we set
\[
H(T,\, x)=\left( \int_I U(x,\,\xi) \sqrt{ \, \frac{f(T,\, \xi)}{\, \xi^2+f(T,\, \xi)\,} \,} \, \tanh \frac{\,\sqrt{\,\xi^2+f(T,\, \xi)\,}\,}{2T}\, d\xi \right)^2
\]
at all $(T,\, x) \in D$. Then
\begin{eqnarray*}
\left| Af(T,\,x)-H(T,\, x) \right| &\leq& \left| Af(T,\,x)-Af_n(T,\, x) \right|+\left| Af_n(T,\,x)-H(T,\, x) \right| \\
&\leq& \left\| Af-Af_n \right\|+\left| Af_n(T,\,x)-H(T,\, x) \right|.
\end{eqnarray*}
By the proof of Lemma \ref{lm:overlineW},
\[
\left\| Af-Af_n \right\| \to 0
\]
as $n \to \infty$. On the other hand, a proof similar to that of Lemma \ref{lm:acontinuous} gives
\[
\left| Af_n(T,\,x)-H(T,\, x) \right| \leq \left( 2 \, \frac{ \, U_2^2 \,}{U_1^2} \sqrt{a}+
 \frac{ \, U_2^2 \, \Delta_2(0)^2 \,}{U_1 \, \varepsilon^2} \right) \, \| f_n-f \| \to 0
\]
as $n \to \infty$. The result thus follows.
\end{proof}

A straightforward calculation gives the following.

\begin{lemma}\label{lm:rlc} \quad $\displaystyle{ A: \, \overline{W} \to \overline{W}}$ is continuous. Moreover, the set $A\overline{W}$ is uniformly bounded and equicontinuous, and hence the set $A\overline{W}$ is relatively compact.
\end{lemma}

Lemma \ref{lm:rlc} immediately implies the following.

\begin{lemma}\label{lm:acompact}
The operator $A:\,  \overline{W} \to \overline{W}$ is compact. Therefore, the operator $A:\,  \overline{W} \to \overline{W}$ has a unique fixed point $f_0 \in \overline{W}$, i.e., $\displaystyle{ f_0=Af_0 }$.
\end{lemma}

\begin{proof}
Applying the Schauder fixed-point theorem gives that the operator $A:\,  \overline{W} \to \overline{W}$ has at least one fixed point $f_0 \in \overline{W}$. A proof similar  to that of \cite[Lemma 3.10]{watanabe-one} gives the uniqueness of $f_0 \in \overline{W}$. 
\end{proof}

Our proof of Theorem \ref{thm:main} is now complete.

\bigskip

In order to give a proof of Theorem \ref{thm:maintwo}, we need to deal with the thermodynamic potential $\Omega$ and differentiate it with respect to the temperature $T$ twice, as mentioned before. Note that the thermodynamic potential $\Omega$ has the fixed point $f_0 \in \overline{W}$ given by Theorem \ref{thm:main} in its form, not the solution $\sqrt{ f_0 }$ to the BCS-Bogoliubov gap equation. Suppose that the fixed point $f_0$ is an element of the subset $W$. It then follows immediately from Theorem \ref{thm:main} that $f_0 \in C^2(D)$. Hence the thermodynamic potential $\Omega$ with the fixed point $f_0$ satisfies all the conditions in the operator-theoretical definition of the second-order phase transition (see \cite[Definition 1.10]{watanabe-five}). We thus apply a proof similar to that of \cite[Theorem 2.4]{watanabe-five} to have Theorem \ref{thm:maintwo}.

Suppose that the fixed point $f_0$ is an accumulating point of the subset $W$. We then replace the fixed point $f_0 \in \overline{W} \setminus W$ in the form of the thermodynamic potential $\Omega$ by a suitably chosen element of $f \in W$ since the fixed point $f_0$ is an accumulating point of the subset $W$. Thanks to Theorem \ref{thm:main}, we find that the suitably chosen element $f$ is in $C^2(D)$. Then we can differentiate the suitably chosen element $f$ with respect to the temperature $T$ twice. Therefore, once we replace the fixed point $f_0 \in \overline{W} \setminus W$ in the form of the thermodynamic potential $\Omega$ by a suitably chosen element of $f \in W$, we can again show that the thermodynamic potential $\Omega$ with this $f \in W$ satisfies all the conditions in the operator-theoretical definition of the second-order phase transition. We can again apply a proof similar to that of \cite[Theorem 2.4]{watanabe-five} to have Theorem \ref{thm:maintwo}. This proves Theorem \ref{thm:maintwo}.

%%%%%%%%%%%%%%%%%%%%

\noindent \textbf{Author contributions}

Shuji Watanabe wrote the main manuscript text and reviewed the manuscript.

\noindent \textbf{Funding}

This work was supported in part by JSPS Grant-in-Aid for Scientific Research (C) KAKENHI Grant Number JP21K03346.

\noindent \textbf{Competing interests}

The author declares no competing interests.

\end{document}